\documentclass[twocolumn]{aastex631}

\usepackage{silence} 
\usepackage{times}
\usepackage{enumitem}
\usepackage{graphicx}
\usepackage{amsmath}
\usepackage{bm}
\usepackage{graphics}
\usepackage{url}
\usepackage{epsfig}
\usepackage{wrapfig}
\usepackage{datetime}
\usepackage{ulem}
\usepackage{verbatim}
\usepackage{float}
\usepackage{lastpage}
\usepackage{enumitem}
\usepackage{amssymb}
\usepackage{natbib}

\providecommand{\apjl}{ApJ}

\providecommand{\apjs}{ApJS}
\providecommand{\aap}{A\&A}

\def\simlt{\mathrel{\hbox{\rlap{\hbox{\lower4pt\hbox{$\sim$}}}\hbox{$<$}}}}
\def\simgt{\mathrel{\hbox{\rlap{\hbox{\lower4pt\hbox{$\sim$}}}\hbox{$>$}}}}

\def\hst{{\it HST}}
\def\spitzer{{\it Spitzer}}

\def\I{\,\textsc{i}}

\def\V{\,\textsc{v}}

\def\arcsec{$^{\,\prime\prime}$}

\def\arcdeg{\degr}   

\shorttitle{SN\,2023ixf Progenitor System}
\shortauthors{Kilpatrick et al.}

\begin{document}

\title{SN\,2023ixf in Messier 101: A Variable Red Supergiant as the Progenitor Candidate to a Type II Supernova}

\correspondingauthor{Charles~D.~Kilpatrick}
\email{ckilpatrick@northwestern.edu}

\def\northwestern{Center for Interdisciplinary Exploration and Research in Astrophysics (CIERA) and Department of Physics and Astronomy,\\ Northwestern University, Evanston, IL 60208, USA}
\def\jhu{Department of Physics and Astronomy, The Johns Hopkins University, Baltimore, MD 21218, USA}
\def\psu{Department of Astronomy and Astrophysics, The Pennsylvania State University, Davey Lab, State College, Pa 16802, USA}
\def\ucb{Department of Astronomy and Astrophysics, University of California, Berkeley, CA 94720, USA}
\def\ucsc{Department of Astronomy and Astrophysics, University of California, Santa Cruz, CA 95064, USA}
\def\ifa{Institute for Astronomy, University of Hawaii, 2680 Woodlawn Drive, Honolulu, HI 96822, USA}
\def\dark{DARK, Niels Bohr Institute, University of Copenhagen, Jagtvej 128, 2200 Copenhagen, Denmark}
\def\oxford{Department of Physics, University of Oxford, Denys Wilkinson Building, Keble Road, Oxford OX1 3RH, UK}
\def\icds{Institute for Computational \& Data Sciences, The Pennsylvania State University, University Park, PA, USA}
\def\qub{Astrophysics Research Centre, School of Mathematics and Physics, Queen’s University Belfast, Belfast BT7 1NN, UK}
\def\igc{Institute for Gravitation and the Cosmos, The Pennsylvania State University, University Park, PA 16802, USA}
\def\gemini{Gemini Observatory, NSF’s NOIRLab, 670 N. A’ohoku Place, Hilo, HI 96720, USA}
\def\carnegie{The Observatories of the Carnegie Institution for Science, 813 Santa Barbara St., Pasadena, CA 91101, USA}
\def\uiuc{Department of Astronomy, University of Illinois at Urbana-Champaign, 1002 W. Green St., IL 61801, USA}
\def\cas{Center for Astrophysical Surveys, National Center for Supercomputing Applications, Urbana, IL, 61801, USA}
\def\cambridge{Institute of Astronomy and Kavli Institute for Cosmology, Madingley Road, Cambridge, CB3 0HA, UK}
\def\toronto{David A. Dunlap Department of Astronomy and Astrophysics, University of Toronto, 50 St. George Street, Toronto, Ontario, M5S 3H4 Canada}

\author[0000-0002-5740-7747]{Charles~D.~Kilpatrick}
\affil{\northwestern}

\author[0000-0002-2445-5275]{Ryan~J.~Foley}
\affiliation{\ucsc}

\author[0000-0002-3934-2644]{Wynn~V.~Jacobson-Gal\'{a}n}
\affil{\ucb}

\author[0000-0001-6806-0673]{Anthony~L.~Piro}
\affil{\carnegie}

\author[0000-0002-8229-1731]{Stephen~J.~Smartt}
\affil{\qub}
\affil{\oxford}

\author[0000-0001-7081-0082]{Maria~R.~Drout}
\affil{\toronto}

\author[0000-0003-4906-8447]{Alexander~Gagliano}
\affil{\uiuc}
\affil{\cas}
\affil{NSF Graduate Research Fellow}

\author[0000-0002-8526-3963]{Christa~Gall}
\affiliation{\dark}

\author[0000-0002-4571-2306]{Jens~Hjorth}
\affil{\dark}

\author[0000-0002-6230-0151]{David~O.~Jones}
\affil{\gemini}

\author[0000-0001-9846-4417]{Kaisey~S.~Mandel}
\affil{\cambridge}

\author[0000-0003-4768-7586]{Raffaella~Margutti}
\affil{\ucb}

\author[0000-0003-2558-3102]{Enrico Ramirez-Ruiz}
\affil{\ucsc}

\author[0000-0003-4175-4960]{Conor~L.~Ransome}
\affil{\psu}

\author[0000-0002-5814-4061]{V.~Ashley~Villar}
\affil{\psu}
\affil{\icds}
\affil{\igc}

\author[0000-0003-4263-2228]{David~A.~Coulter}
\affil{\ucsc}

\author[0000-0003-1015-5367]{Hua~Gao}
\affil{\ifa}

\author[0000-0002-4513-3849]{David~Jacob~Matthews}
\affil{\ucb}

\author[0000-0002-5748-4558]{Kirsty~Taggart}
\affil{\ucsc}

\author[0000-0002-0632-8897]{Yossef~Zenati}
\affil{\jhu}
\affil{ISEF International Fellowship}

\begin{abstract}

We present pre-explosion optical and infrared (IR) imaging at the site of the type\,II supernova (SN\,II) 2023ixf in Messier 101 at 6.9~Mpc.  We astrometrically registered a ground-based image of SN\,2023ixf to archival {\it Hubble Space Telescope} ({\it HST}), {\it Spitzer Space Telescope} ({\it Spitzer}), and ground-based near-IR images.  A single point source is detected at a position consistent with the SN at wavelengths ranging from {\it HST} $R$-band to {\it Spitzer} 4.5~$\mu$m.  Fitting to blackbody and red supergiant (RSG) spectral-energy distributions (SEDs), we find that the source is anomalously cool with a significant mid-IR excess.  We interpret this SED as reprocessed emission in a 8600~$R_{\odot}$ circumstellar shell of dusty material with a mass $\sim$5$\times10^{-5}~M_{\odot}$ surrounding a $\log(L/L_{\odot})=4.74\pm0.07$ and $T_{\rm eff}=3920\substack{+200\\-160}$~K RSG.  This luminosity is consistent with RSG models of initial mass 11~$M_{\odot}$, depending on assumptions of rotation and overshooting.  In addition, the counterpart was significantly variable in pre-explosion {\it Spitzer} 3.6~$\mu$m and 4.5~$\mu$m imaging, exhibiting $\sim$70\% variability in both bands correlated across 9~yr and 29 epochs of imaging.  The variations appear to have a timescale of 2.8~yr, which is consistent with $\kappa$-mechanism pulsations observed in RSGs, albeit with a much larger amplitude than RSGs such as $\alpha$ Orionis (Betelgeuse).

\end{abstract}


\section{Introduction}\label{sec:introduction}

All hydrogen-rich supernovae (SN\,II) with directly identified progenitor stars have been interpreted to come from systems with initial mass $<$20~$M_{\odot}$ \citep{Smartt15}.  With the exception of the blue supergiant progenitor of the peculiar SN\,II 1987A \citep{Hillebrandt87,arnett+89} and luminous blue variable (LBV) progenitor stars to SN\,IIn \citep[e.g.,][]{galyam09}, all of these systems are red supergiants (RSGs).  These stars have massive, extended, hydrogen envelopes and make up the majority of directly-identified progenitor stars to core-collapse SNe \citep[SNe~2003gd, 2004A, 2004et, 2005cs, 2006my, 2008bk, 2009hd, 2009kr, 2009md, 2012A, 2012aw, 2012ec, 2016cok, 2017eaw, 2018aoq, 2020jfo, 2022acko;][]{Smartt04,Maund09a,Fraser10,Crockett11,Elias-Rosa11,Fraser11,Kochanek12,Maund13,Tomasella13,Fraser14,Maund14,Kochanek17,Kilpatrick18:17eaw,ONeill19,Rui19,vanDyk19,Sollerman21,vanDyk23a,vanDyk23b}.  The lack of $>$20~$M_{\odot}$ RSG progenitor stars to SN\,II despite the fact that they make up $\approx$15\% of RSGs following a Salpeter initial mass function and RSGs with $\log(L/L_{\odot})>5.2$ are observed in the LMC, M31, and M33 \citep{Drout12,Neugent20,Neugent20a,Neugent21} has been noted as the ``red supergiant problem'' \citep[][although see also \citealt{Davies18}]{Smartt09}.

Theoretically, massive RSGs are predicted to have compact oxygen cores, and many of them may collapse directly to black holes as ``failed SNe,'' leading to a paucity of high-mass counterparts to SN\,II \citep{Sukhbold16}.  This scenario broadly agrees with the light curves and nucleosynthetic yields of SN\,II, which also favor lower mass progenitor stars \citep{Brown13,Muller17,Morozova18} as well as direct evidence for a high-mass RSG in NGC\,6946 whose optical counterpart disappeared \citep[][see also \citealt{Neustadt21b} and \citealt{Byrne22}]{adams+16}.  This source also left behind a weak infrared (IR) transient consistent with expectations for mass ejection in failed SNe \citep{Lovegrove13,Piro13,Fernandez18}.  Long time baseline follow up of nearby galaxies with deep, high-resolution imaging can constrain the fraction of disappearing stars, such as the estimate by the ``Survey for Nothing'' that $\approx16\%$ of massive stars produce failed SNe in \citet{Neustadt21}, close to the value required by an upper mass threshold for successful explosions of $>$20~$M_{\odot}$.  Simultaneously constraining the fraction and mass distribution of failed SN and SN\,II progenitor stars is therefore a powerful tool for probing massive star structure and the latest stages of stellar evolution.

SN\,II progenitor stars also exhibit a wide range of circumstellar densities in their immediate vicinity ($<$10$^{15}$~cm) as implied by flash spectroscopy \citep{gal-yam+14,khazov+15,Yaron17,terreran+2022,Tinyanont22}, early photometric evolution \citep{Morozova17,Morozova18}, as well as evidence for pre-explosion variability and eruptions \citep{Kilpatrick18:17eaw,Jacobson-Galan22}.  These features may have significant implications for the interpretation of their pre-explosion counterparts in the absence of multi-band, multi-epoch imaging.  For example, the vast majority of SN\,II pre-explosion counterparts are identified in F814W imaging from the {\it Hubble Space Telescope} \citep[\hst, e.g., in][]{Smartt09,Davies18}.  This filter is blueward of the peak of RSG spectral-energy distributions (SEDs) and may be significantly impacted by circumstellar extinction in the presence of a dusty shell.  Moreover, many RSGs exhibit well-known modes of variability \citep{Stothers69,Jurcevic00,Guo02,Yang11,Soraisam18} that may become even more extreme as they approach core collapse \citep{Yoon10,fuller+17,Davies22}, although SN\,2016cok is a counter-example whose progenitor star exhibited very little variability \citep{Kochanek17}.  Without multi-epoch imaging in which their average luminosities can be estimated, interpretation of photometry for SN\,II pre-explosion counterparts is complicated by large systematic uncertainties.

Here we present pre-explosion imaging to the nearby SN\,II 2023ixf discovered in Messier 101 (M101) on 19 May 2023 \citep{Itagaki23}.  These data cover ultraviolet to mid-IR bands from 1999--2019.  We demonstrate that there is a single credible progenitor candidate to SN\,2023ixf and estimate its luminosity, temperature, and initial stellar mass as well as its variability and total circumstellar material (CSM) inferred from a significant mid-IR excess.  We find it was significantly variable in the mid-IR and compare that timescale with well-observed RSGs.  We summarize the total data set and our reduction procedure in Section~\ref{sec:observations} and analysis and modeling of those data in Section~\ref{sec:progenitor}.  Finally, we discuss the broader implications of this progenitor candidate in Section~\ref{sec:discussion} and conclude in Section~\ref{sec:conclusions}.

We assume a line-of-sight extinction through the Milky Way of $A_{V}=0.025$~mag (assuming $R_{V}=3.1$, this is $E(B-V)=0.008$~mag) from \citet{Schlafly11}.  We also adopt the latest Cepheid distance to M101 of 6.85$\pm$0.15~Mpc from \citet{riess22}.  
Finally, throughout this paper we reference \citet{Jacobson-Galan23}, who demonstrate that SN\,2023ixf appears to be a normal type II SN with broad lines of hydrogen.   We also assume a host reddening to SN\,2023ixf of $E(B-V)=0.033$~mag from \citet{Jacobson-Galan23}, derived from Na\I~D line absorption in optical spectra of this event.  Given the small value for this line-of-sight reddening, it does not significantly impact our results and we adopt a total-to-selective extinction ratio in the host galaxy of $R_{V}=3.1$ (implying $A_{V}=0.10$~mag), however we acknowledge that this could range from $R_{V}=2$--6 (implying $A_{V}=0.07$--0.20~mag).

\section{Observations of SN~2023\lowercase{ixf} and Its Progenitor Candidate}\label{sec:observations}

\subsection{{\it Hubble Space Telescope}}\label{sec:hst}

\begin{deluxetable}{ccccc}
\tablecaption{\hst\ Photometry of the SN\,2023ixf Progenitor Candidate}
\tablehead{
\colhead{MJD} &
\colhead{Instrument} &
\colhead{Filter} & 
\colhead{$m$} &
\colhead{$\sigma_m$} \\
&
&
&
\colhead{(mag)} &
\colhead{(mag)}
}
\startdata
51260.9786 & WFPC2     & F656N & $>$23.433 & -- \\
51261.0390 & WFPC2     & F675W & 26.422 & 0.230 \\
51261.1120 & WFPC2     & F547M & $>$26.273 & -- \\
51345.9897 & WFPC2     & F656N & $>$23.776 & -- \\
51346.0529 & WFPC2     & F547M & $>$26.416 & -- \\
52593.9933 & ACS/WFC   & F435W & $>$27.393 & -- \\
52594.0096 & ACS/WFC   & F555W & $>$27.099 & -- \\
52594.0215 & ACS/WFC   & F814W & 24.881 & 0.059 \\
52878.3224 & WFPC2     & F336W & $>$27.025 & -- \\
53045.0069 & ACS/WFC   & F658N & 25.332 & 0.284 \\
56735.8571 & WFC3/UVIS & F673N & $>$24.629 & -- \\
58207.4384 & ACS/WFC   & F658N & $>$25.488 & -- \\
58207.4561 & ACS/WFC   & F435W & $>$27.799 & --
\enddata
\tablecomments{All magnitudes are in the AB system.}\label{tab:hst}
\end{deluxetable}

The site of SN\,2023ixf was observed with the {\it Hubble Space Telescope}/WFPC2, ACS, and WFC3 over seven epochs from 23 March 1999 to 30 March 2018, or 24.2 to 5.1~years before discovery (\autoref{tab:hst}).  Following methods described in \citet{Kilpatrick21b} and \citet{Kilpatrick21}, we used a custom {\tt python}-based pipeline {\tt hst123}\footnote{\url{https://github.com/charliekilpatrick/hst123}} to download, align, and drizzle all {\it HST} imaging \citep[for details, see][]{drizzlepac}, and perform photometry in {\tt dolphot} \citep{dolphot}.  We used recommended {\tt dolphot} settings for each imager as described in the respective manual\footnote{\url{americano.dolphinsim.com/dolphot}}.

The final stacked imaging of M101 observed in 2002 by ACS is shown in \autoref{fig:astrometry} as a RGB image (F814W, F555W, F435W).  We also show each image in which we obtain a detection of a counterpart at the explosion site of SN\,2023ixf, which includes WFPC2 F675W and ACS F658N imaging.  In addition, we have deep constraints in bluer bands F336, F435W, and F555W, which we consider in the context of a binary companion below.

Within 0.2\arcsec\ of the reported position of SN\,2023ixf, there are two sources detected in F814W.  This is clearly seen in Fig.~\ref{fig:astrometry} where a counterpart is located at the site of SN\,2023ixf in the ACS imaging and appears as a blended source in the F814W panel.  The brighter source has $m_{\rm F814W}=24.881\pm0.059$~mag\footnote{All photometry reported throughout this paper is on the AB magnitude system.} that we refer to as ``Source A'' and is blended with the fainter ``Source B'' approximately 0.1\arcsec\ (2.0 ACS/WFC pixels) to the northeast with $m_{\rm F814W}=25.955\pm0.125$~mag.  Below we consider which, if either, of these sources may be the pre-explosion counterpart to SN\,2023ixf and the extent to which any blended emission from other sources may contaminate photometry of that source in other bands with poorer resolution.

\begin{figure*}
    \centering
    \includegraphics[width=\textwidth]{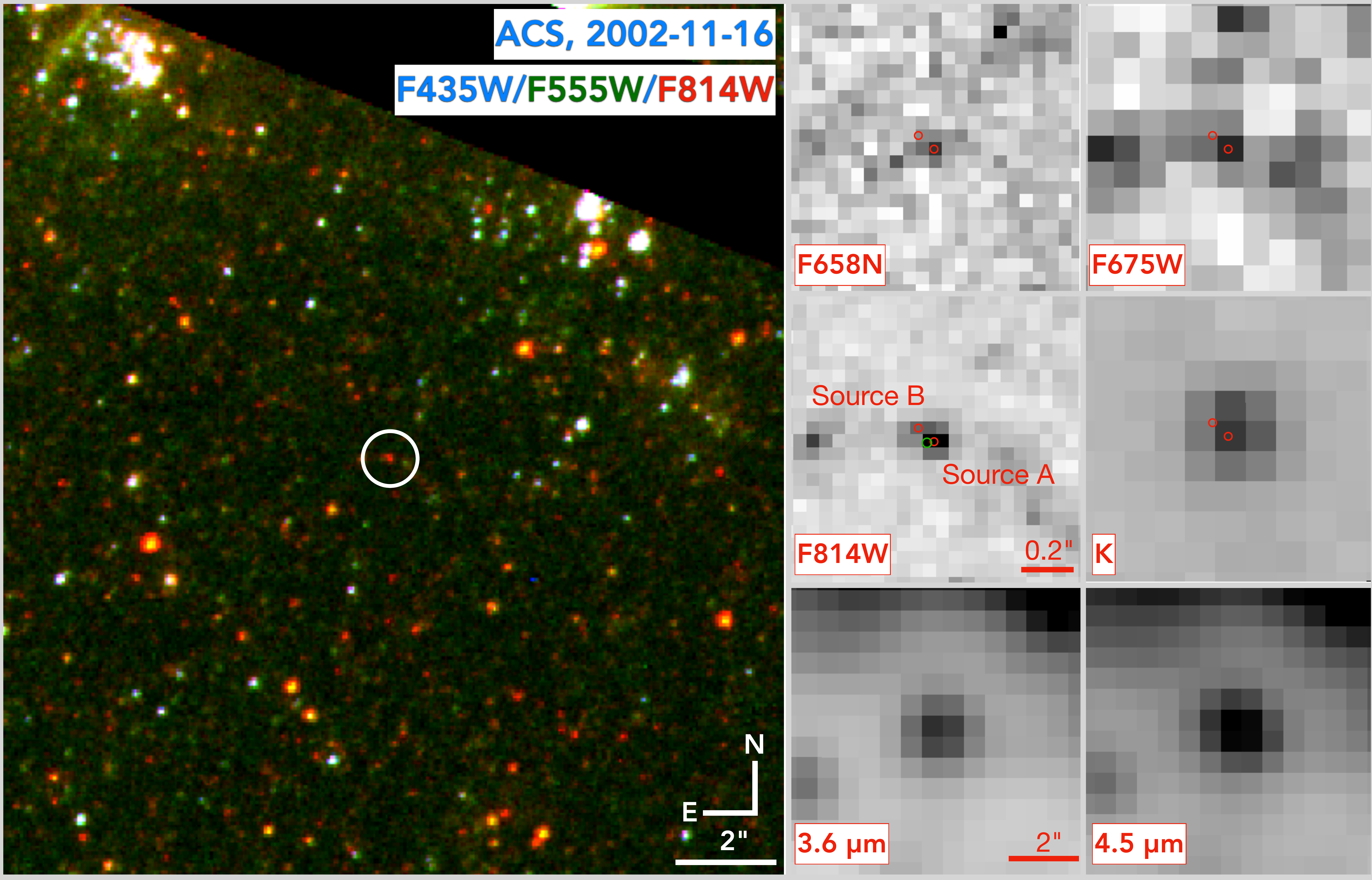}
    \caption{({\it Left}): A 15.4\arcsec$\times$13.1\arcsec\ cutout of {\it HST}/ACS imaging toward M101 in F435W (blue), F555W (green), and F814W (red).  We show the approximate explosion site of SN\,2023ixf as a 1\arcsec\ diameter white circle.  ({\it Right panels}): Panels showing pre-explosion F658N, F675W, F658N, Gemini/NIRI $K$-band, and {\it Spitzer} Channel 1 (3.6~$\mu$m) and 2 (4.5~$\mu$m) imaging where we detect a counterpart at the site of SN\,2023ixf.  The \hst\ and $K$-band images are on the same scale, while the {\it Spitzer} 3.6 and 4.5~$\mu$m show a zoomed out scale to highlight the location of the counterpart.  We show red circles corresponding to the locations of Sources A and B, close to the site of SN\,2023ixf as discussed in Section~\ref{sec:hst}.  We also note the position of the SN (with the approximate systematic uncertainty 0.04\arcsec) with a green circle in the F814W panel (see Section~\ref{sec:alignment}).}
    \label{fig:astrometry}
\end{figure*}

\begin{figure}
    \includegraphics[width=0.49\textwidth]{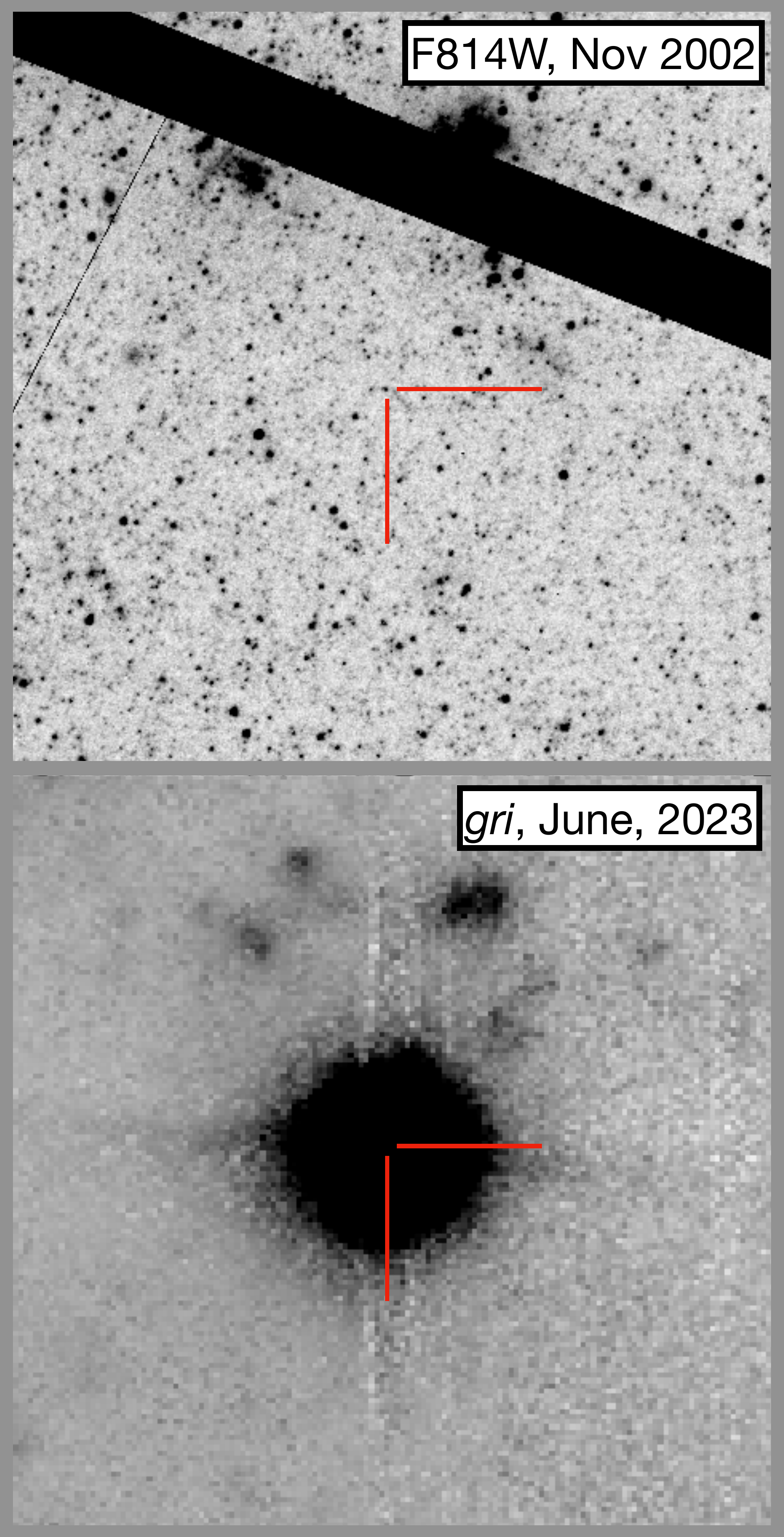}
    \caption{A part of {\it HST}/ACS F814W imaging (top) of M101 from 16 November, 2002 compared with a $gri$ image from Gemini-N/GMOS (bottom) of the same field from June, 2023 showing the location of SN\,2023ixf.  We identify a single counterpart at the position of SN\,2023ixf in the ACS image, discussed in detail in Section~\ref{sec:alignment}.}\label{fig:alignment}
\end{figure}

\subsection{{\it Spitzer Space Telescope}/IRAC}\label{sec:spitzer}

The site of SN\,2023ixf was observed over 31 epochs with the {\it Spitzer Space Telescope} Infrared Array Camera (\spitzer/IRAC) from 8 March 2004 to 25 October 2019, roughly 19.2 to 3.6~yr prior to discovery.  We obtained all such imaging for the cold and warm \spitzer\ mission from the Spitzer Heritage Archive\footnote{\url{https://sha.ipac.caltech.edu/}}.  Following methods described in \citet{Fox21} and \citet{Rubin21}, we applied a forward-modeling approach to estimate the Channel 1 and 2 (3.6 and 4.5~$\mu$m, respectively) fluxes of point-like emission near the site of SN\,2023ixf.  After stacking and mosaicking the individual epochs in {\tt MOPEX} \citep{Makovoz05}, we identified a single point source within 2\arcsec\ of the site of SN\,2023ixf, originally reported in \citet{Szalai23} and detected in all Channel 1 and 2 frames.  We estimated the total flux of this source in the individual basic calibrated data ({\tt cbcd}) frames across each epoch using realistic point-spread functions (PSFs) for the warm \spitzer\ mission and a {\tt python}-based forward modeling code\footnote{\url{https://github.com/charliekilpatrick/forwardmodel}}.  This photometry is given in Table~\ref{tab:spitzer}.  We also provide the average photometry, which we use in our modeling below with added uncertainty accounting for the individual error bars and standard deviation across all epochs.

To validate our \spitzer\ photometry, we used our photometry code to analyze a sequence of 8--10 stars across every image.  We looked for variability that may arise from instrumental effects or the position of the spacecraft at the time of observation.  Our photometry indicates that all stars exhibit very little variability (i.e., at the $<$5\% level) compared with photometry of the SN\,2023ixf counterpart across each epoch.  We conclude that variations in the counterpart are intrinsic to that source as opposed to systematic effects.

In addition, there were two epochs of Channels 3 and 4 (5.4 and 8.0~$\mu$m, respectively) data obtained at the site of SN\,2023ixf from the cold {\it Spitzer} mission.  We do not detect any significant source of emission in either epoch and instead place a forced circular aperture with a size of 3.0 and 3.4\arcsec, or approximately 2$\times$ the full-width at half-maximum of the \spitzer\ point-response function, at the site of SN\,2023ixf.  From these data, we estimate a 3$\sigma$ upper limit on the presence of any emission, which is given for both epochs in Table~\ref{tab:spitzer}.  We also estimate a flux-weighted average and standard deviation of all Channel 1 and 2 photometry as well as limiting fluxes for the stacked Channel 3 and 4 imaging obtained using the same method as the individual epochs, which are all given as the last four rows in Table~\ref{tab:spitzer}.

Finally, we consider the possibility that the {\it Spitzer} counterpart is a blend of Sources A and B, or other sources not visible in the \hst\ bands.  If all sources have similar optical-IR colors, then deblending Source A and B in the {\it Spitzer} frames could reduce the flux of the true counterpart by 27\%, comparable in magnitude to our error bars on the average values.  Additional follow-up observations with {\it JWST} at late times would resolve any emission at the scale of the Source A and B separation, enabling a cleaner subtraction of any blended emission.  For the analysis below, we assume that the {\it Spitzer} flux is entirely dominated by the SN\,2023ixf counterpart.

\startlongtable
\begin{deluxetable}{ccccc}
\tablecaption{\spitzer/IRAC Photometry of the SN\,2023ixf Progenitor Candidate}
\tablehead{
\colhead{MJD} &
\colhead{Band} &
\colhead{Flux} &
\colhead{Uncertainty} \\
&
&
\colhead{($\mu$Jy)} &
\colhead{($\mu$Jy)}
}
\startdata
53072.0903 & Ch1 & 29.80 & 2.22 \\
53072.0903 & Ch2 & 28.97 & 2.74 \\
53072.0903 & Ch3 & $<$26.78 & -- \\
53072.0903 & Ch4 & $<$30.60 & -- \\
53072.4901 & Ch1 & 29.01 & 2.88 \\
53072.4901 & Ch2 & 26.55 & 3.41 \\
53072.4901 & Ch3 & $<$27.00 & -- \\
53072.4901 & Ch4 & $<$30.72 & -- \\
55960.7226 & Ch1 & 17.82 & 2.50 \\
55980.9934 & Ch1 & 17.50 & 2.44 \\
56165.0117 & Ch2 & 17.85 & 2.53 \\
56337.0654 & Ch1 & 16.96 & 2.77 \\
56348.1056 & Ch1 & 19.52 & 2.48 \\
56516.3523 & Ch2 & 19.01 & 2.55 \\
56742.8361 & Ch1 & 29.45 & 2.30 \\
56742.8361 & Ch2 & 32.09 & 1.97 \\
56771.8253 & Ch1 & 29.94 & 2.31 \\
56771.8253 & Ch2 & 31.37 & 1.75 \\
56902.0136 & Ch1 & 24.27 & 3.31 \\
56902.0136 & Ch2 & 26.52 & 3.24 \\
57136.6924 & Ch1 & 17.71 & 2.20 \\
57136.6924 & Ch2 & 22.66 & 1.83 \\
57144.0597 & Ch1 & 19.24 & 2.01 \\
57144.0597 & Ch2 & 20.47 & 1.86 \\
57150.1719 & Ch1 & 21.23 & 1.92 \\
57150.1719 & Ch2 & 19.67 & 1.72 \\
57163.7124 & Ch1 & 21.37 & 1.77 \\
57163.7124 & Ch2 & 25.15 & 1.51 \\
57191.8234 & Ch1 & 14.30 & 2.36 \\
57191.8234 & Ch2 & 19.11 & 2.03 \\
57220.7940 & Ch1 & 15.19 & 2.69 \\
57220.7940 & Ch2 & 18.19 & 2.32 \\
57247.8227 & Ch1 & 15.53 & 2.88 \\
57247.8227 & Ch2 & 14.94 & 2.37 \\
57486.8506 & Ch1 & 21.44 & 2.18 \\
57486.8506 & Ch2 & 24.44 & 1.64 \\
57843.9334 & Ch1 & 27.66 & 2.27 \\
57843.9334 & Ch2 & 30.23 & 2.00 \\
57926.9005 & Ch2 & 26.65 & 1.98 \\
58009.6705 & Ch1 & 22.82 & 2.73 \\
58009.6705 & Ch2 & 20.87 & 2.66 \\
58232.9534 & Ch1 & 21.63 & 2.13 \\
58232.9534 & Ch2 & 20.51 & 1.98 \\
58292.8692 & Ch1 & 17.46 & 1.37 \\
58292.8692 & Ch2 & 21.34 & 1.68 \\
58380.2202 & Ch1 & 18.29 & 2.84 \\
58380.2202 & Ch2 & 17.32 & 2.53 \\
58572.0773 & Ch1 & 22.39 & 2.54 \\
58572.0773 & Ch2 & 26.20 & 2.03 \\
58614.3896 & Ch1 & 22.33 & 2.44 \\
58614.3896 & Ch2 & 27.38 & 1.85 \\
58655.6803 & Ch1 & 25.47 & 2.06 \\
58655.6803 & Ch2 & 28.60 & 1.81 \\
58697.4982 & Ch1 & 28.49 & 2.53 \\
58697.4982 & Ch2 & 27.30 & 2.07 \\
58740.0119 & Ch1 & 22.42 & 3.17 \\
58740.0119 & Ch2 & 23.75 & 2.62 \\
58781.3131 & Ch1 & 28.26 & 3.17 \\
58781.3131 & Ch2 & 32.80 & 2.41 \\
\hline
\multicolumn{4}{c}{Average \spitzer/IRAC Photometry} \\
\hline
--         & Ch1 & 22.13 & 4.78 \\
--         & Ch2 & 23.99 & 4.87 \\
--         & Ch3 & $<$21.64 & -- \\
--         & Ch4 & $<$24.15 & -- \\
\enddata
\tablecomments{See \autoref{sec:spitzer}.}\label{tab:spitzer}
\end{deluxetable}

\subsection{Ground-based Infrared Imaging}

The NEWFIRM infrared camera \citep{Autry03} observed M101 in $JHK_{s}$ bands from 29 June to 1 July 2010.  We obtained these data as reduced and sky-subtracted image frames from the NOIRLab data archive\footnote{\url{https://astroarchive.noirlab.edu/}}.  Stacking the frames for each band in {\tt swarp} \citep{swarp} using flux scaling derived from the calibration in their image headers, we recalibrated the final coadded image using {\tt DoPhot} PSF photometry \citep{schechter+93} and 2MASS $JHK_{s}$ photometric calibrators in the same image frame as the NEWFIRM images \citep{Skrutskie06}.  We detect a point-like counterpart within 2\arcsec\ of the site of SN\,2023ixf in the $K_{s}$ image, which is also the deepest NEWFIRM image overall.  In summary, we find that this source has $m_{K_{s}}=20.74\pm0.15$~mag, $m_{H}>20.36$~mag and $m_{J}>20.18$~mag.

The site of SN\,2023ixf was also observed by the Gemini-North telescope with the Near-Infrared Imager (NIRI) on 18 April 2010 using a $K$-band continuum filter and 51$\times$50~s exposures.  We processed all such imaging using {\tt pyraf}-based methods from the Gemini {\tt IRAF} library \citep{Cooke05} developed for NIRI, including dark-frame subtraction, flat-fielding, sky subtraction, and optimal alignment and image coadding.  We performed photometry following the same methods described above for the NEWFIRM imaging, however there were only two 2MASS $K_{s}$ standard stars in the NIRI image.  Therefore, we use photometry of all objects other than the SN\,2023ixf counterpart that are classified as bright point sources (Object type 1) by {\tt DoPhot} from the calibrated NEWFIRM $K_{s}$ image to calibrate the NIRI image.  Similar to the NEWFIRM imaging, there is a single point-like source within 2\arcsec\ of the SN\,2023ixf position, shown in \autoref{fig:astrometry}.  We find that this source has $m_{K}=20.72\pm0.08$~mag.

\subsection{GMOS Imaging of SN~2023ixf}

Gemini-N/GMOS obtained a series of 10$\times$1~s images and 3$\times$60~s images in $gri$ bands on 3 and 5 June 2023.  We obtained all such imaging from the Gemini Data Archive\footnote{\url{https://archive.gemini.edu/}}, and following standard procedures in {\tt astropy}, we removed the bias from these frames using the overscan correction.  We then calibrated each frame using {\tt DoPhot} photometry \citep{schechter+93} and Pan-STARRS standard stars in each frame of the GMOS images \citep{flewelling+16}.  To obtain the deepest possible image of the field surrounding SN\,2023ixf without saturating the SN position, we masked saturated pixels from the SN and stacked all images in {\tt swarp} \citep{swarp} into a single $g+r+i$ frame, weighting the individual frames by the inverse variance of the sky pixels across the individual bands.  The final image centered on the position of SN\,2023ixf is shown in Fig.~\ref{fig:alignment}.

\section{The Progenitor Candidate of SN~2023\lowercase{ixf}}\label{sec:progenitor}

\subsection{Aligning Pre- and Post-Explosion Imaging}\label{sec:alignment}

To establish that SN\,2023ixf is associated with a candidate counterpart in pre-explosion imaging, we align our post-explosion GMOS image to the ACS F814W frame and determine astrometrically whether the SN is consistent with coming from any point-like sources.  Although this method can rule out an association between the SN and any pre-explosion counterpart, high-resolution follow up imaging after the SN has faded is needed to establish that any counterpart has disappeared and the association was not a chance coincidence.

We identify 108 common sources between the ACS F814W frame and Gemini $g+r+i$ image frame, which we use to establish a coordinate transformation using the {\tt IRAF} package {\tt ccmap}.  The root-mean square offset from this coordinate transformation solution is $0.02$\arcsec\ in both right ascension and declination.  To determine the systematic uncertainty in our coordinate transformation \citep[and following methods from][]{Kilpatrick21}, we take half of the astrometric calibrator sources from our sample and recalculate the coordinate transformation.  We then estimate the offsets between the remaining stars.  Repeating this process 100 times, we find that the average offset between stars across all trials is $\sim0.03$\arcsec.  In total, we estimate a $0.04$\arcsec\ uncertainty (1.3~pc at the distance of M101) in our alignment between the two frames.

The position of SN\,2023ixf aligns with Source A (Fig.~\ref{fig:alignment}) to within 0.8$\sigma$, while it is 2.4$\sigma$ away from Source B. Thus while SN\,2023ixf could be astrometrically consistent with either source, there is a strong preference for Source A in our analysis.  There is no evidence for a second source in any other image frame that we analyze, and both SN\,2023ixf and Source A are astrometrically consistent with being the same object as the point sources we identified in WFPC2/F675W, ACS/F658N, Gemini/NIRI $K$-band, and {\it Spitzer} Channels 1 and 2 across all epochs.  We provide all photometry of that source in Tables~\ref{tab:hst} and \ref{tab:spitzer}\footnote{All photometry of the SN\,2023ixf progenitor candidate and metadata used in the analysis below is provided in machine-readable format at \url{https://github.com/charliekilpatrick/progenitors/blob/main/sed/data/input/2023ixf.dat}.}.

Finally, we estimate the probability of chance coincidence with Source A by noting that there are 238 sources detected at $>$3$\sigma$ within 3\arcsec\ (100~pc at the distance of M101) of that source.  Therefore, there is a 4\% chance of a single source landing within 1$\sigma$ of the astrometric uncertainty of SN\,2023ixf by chance.  While this is a moderately large probability of chance coincidence, the likelihood could be reduced significantly with high-resolution follow-up imaging and, eventually, by determining whether the candidate counterpart has disappeared with follow-up observations after SN\,2023ixf fades.

\subsection{The Spectral-Energy Distribution of the SN~2023ixf Progenitor System}\label{sec:sed}

\subsubsection{Single Blackbody Fit}

Assuming that the SN\,2023ixf pre-explosion counterpart is dominated by a single SED from its progenitor star and with no variability between each epoch (though see {\it Spitzer} analysis in Section~\ref{sec:variability}), we can model the nature of this source from the ultraviolet to mid-IR.  We initially adopt a simple blackbody spectrum and derive its temperature and luminosity using a Markov Chain Monte Carlo (MCMC) approach in the {\tt python}-based package {\tt emcee}.  Assuming the distance, Milky Way extinction, and host extinction given above, we derive the in-band magnitudes for a blackbody spectrum of a given temperature and luminosity using the filter transmission functions for each space- and ground-based bandpass and using {\tt pysynphot}.  Following methods in \citet{Kilpatrick21}, we fit a blackbody model by sampling the posterior distribution over the range of model parameters, and report their posterior means and standard deviations.

Following this method, we find that the SN\,2023ixf progenitor candidate is consistent with a $\log(L/L_{\odot})=4.73\substack{+0.07\\-0.08}$ and $T_{\rm eff}=1640\pm20$~K blackbody as shown in \autoref{fig:sed}.  This implied photospheric temperature is extremely low, even for the latest M supergiant spectral types \citep[i.e., the coolest RSGs have $T_{\rm eff}=3400$--3500~K;][]{Levesque06,Davies13,Davies18}.  The photospheric radius implied by $T_{\rm eff}=1640$~K ($\approx$2900~$R_{\odot}$) suggests that we are seeing material extended well beyond the envelope of a RSG such as a shell of CSM in the local environment around the SN\,2023ixf progenitor star.  We infer that the photosphere in the IR is dominated by a component of host dust, and instead turn to a multi-component SED fit below.

\subsubsection{MARCS Red Supergiant and Circumstellar Dust Fit}

To fit the counterpart with a more realistic optical to mid-IR SED, we use a combined RSG and dusty CSM spectrum initially presented in \citet{Kilpatrick18:17eaw} and based on {\tt DUSTY} radiative transfer models \citep[from][and see also \citealt{Ivezic97,Ivezic99,Elitzur01}]{Kochanek12}.  This model uses a MARCS RSG spectrum of an arbitrary temperature and luminosity \citep[see][for details]{gustaffson+08}, which is reprocessed through a shell of graphitic dust at a given temperature and mass \citep[i.e., similar to carbon-bearing species around massive RSGs, e.g., in][]{Royer10}.
In general, we fit for the RSG luminosity, stellar temperature, dust temperature, and $V$-band optical depth through the mass of dust.  These assumptions yield a mass, luminosity, and radius for the dust assuming a $r^{-2}$ density profile \citep[see][for more details]{Kochanek12,Kilpatrick18:17eaw}.  Assuming a dust-to-gas ratio and wind speed, we can then derive the total mass of CSM and mass-loss rate, which we give below.

From this model, we find that the effective temperature of the counterpart is more in line with known RSGs at $T_{\rm eff}=3920\substack{+200\\-160}$~K while the overall luminosity remains the same at $\log(L/L_{\odot})=4.74\pm0.07$.  In order to fit the IR excess observed in the \spitzer\ bands, we require a circumstellar shell of dust with a $V$-band optical depth of $\tau_{V}=5.8\pm0.2$ (corresponding to $A_{V}=4.6\pm0.2$~mag) and an effective temperature of $T_{\rm dust}=880\pm40$~K.  This shell would have an effective radius of $8600\substack{+900\\-800}$~$R_{\odot}$ and a total dust mass of $5.0\substack{+1.1\\-0.8}\times10^{-7}$~$M_{\odot}$, or a total mass of $5\times10^{-5}~M_{\odot}$ assuming a dust-to-gas ratio of 0.01 in the circumstellar environment \citep[consistent with ratios in the environments of SNe from][]{fox+10}.  

Assuming it was being produced by a constant wind with a $r^{-2}$ profile with a velocity of $v_{\rm wind}$, the implied mass-loss rate is $\dot{M}/(v_{\rm wind})=1.3\pm0.1\times10^{-6}~M_{\odot}~\text{yr}^{-1}/(50~\text{km~s}^{-1})$.  We assume a velocity of 50~km~s$^{-1}$ for consistency with \citet{Jacobson-Galan23} noting that this value is similar to other RSGs \citep[e.g., $30$--$50$~km~s$^{-1}$ for VY~CMa or NML~Cyg;][]{knapp+82,decin+06}, however high-resolution spectroscopy of unshocked material from early in the evolution of SN\,2023ixf can more precisely constrain this value.  

In the fits above, we do not include the ACS F658N detection despite it being spatially coincident with Source A, as this counterpart is likely dominated by H$\alpha$ emission that we do not include in our model.  However, assuming that the emission in this filter contains H$\alpha$ and continuum emission from a RSG, we estimate that the total H$\alpha$ flux density is 1.3$\times$10$^{-19}$~erg~s$^{-1}$~cm$^{-2}$~\AA$^{-1}$ (corrected for host and Milky Way extinction).  This corresponds to a total H$\alpha$ luminosity of 4.9$\times$10$^{39}$~erg~s$^{-1}$ or 1300~$L_{\odot}$.  This value far exceeds the expectations for H$\alpha$ emission in massive RSGs \citep[e.g., 1~$L_{\odot}$ for VY CMa in][]{Smith01}, and so may be unassociated with the progenitor star.

Assuming that the underlying star implied by our RSG model is a single source with $\log(L/L_{\odot})=4.74\pm0.07$, we consider its initial mass by comparing to MIST \citep{choi+16}, STARS \citep{EldridgeTout04}, Geneva \citep{Ekstrom12}, and KEPLER \citep{woosley+07} models.  All models assume a Solar metallicity, and either a rotating or non-rotating star.  In general, the final luminosity of a model SN progenitor depends on the He core luminosity, which is higher in models including rotation and overshooting.  For a MIST model assuming a star at Solar metallicity, we find an initial mass of $11\pm1~M_{\odot}$.  
The STARS models indicate $M_{\rm ZAMS}=11\pm1$$M_{\odot}$, while the the Geneva rotating models produce a star of similar final luminosity at $M_{\rm ZAMS}=11-12$$M_{\odot}$.  
Finally, the KEPLER (non-rotating) models indicate $M_{\rm ZAMS}=12\pm1$.  
Thus the star could feasibly come from a system ranging from 10--13~$M_{\odot}$. 
Second dredge-up in 6--9$M_{\odot}$ stars can increase the final luminosity  substantially \citep[e.g.,][]{Eldridge+07,Jones+13}, with the stars along the asymptotic giant (AGB) branch ending up more luminous and with cooler photospheric temperatures than their more massive M-type counterparts. While our {\tt DUSTY} SED models favor a higher $T_{\rm eff}$ than those of AGB stars (typically $T_{\rm eff}\sim3100$\,K) it is possible the progenitor is a cooler AGB-type star. Measurements of the nickel  mass created in the explosion and the the oxygen mass ejected (both after 100--200 days) will provide interesting constraints on the core mass and explosion mechanism.

Finally, we examine the consistency of our derived mass-loss rate and initial mass with prescriptions from \citet{Beasor19}.  Applying their luminosity-dependent mass-loss rates, we find that for RSGs of $\log(L/L_{\odot})=4.74$ they predict  $\dot{M}=0.7$--4.2$\times$10$^{-6}~M_{\odot}~\text{yr}^{-1}$,\footnote{See eqn. 3 and table 4 in \citet{Beasor19}.} which is in close agreement with our inferred value.  Similarly, applying their initial mass (for 10--12~$M_{\odot}$) and luminosity-dependent parameterization\footnote{See eqn. (4) in \citet{Beasor19}.}, we derive 0.4--1.1$\times$10$^{-6}~M_{\odot}~\text{yr}^{-1}$.  These values are close to our derived mass-loss rate of 1.3$\times$10$^{-6}~M_{\odot}~\text{yr}^{-1}$, especially considering the uncertain wind velocity and uncertainties in model fitting parameters.

\subsubsection{Constraints on a Binary Companion from Optical Limits}

We also consider the possibility that the progenitor star evolved in a binary and exploded as the primary star in that system.  Comparing our photometry to BPASS v2.2.1 binary star models \citep{eldridge+17}, we examine all systems for which the combined flux from the primary and secondary at the time the primary explodes is fainter than our limiting magnitudes.  We emphasize that these models do not include circumstellar extinction or predictions for the mid-IR luminosity, and we only use the limits from our bluer bands where we predict the primary star to be faint in order to constrain the presence of a companion star.  

Although we examine all bands contained in the BPASS models for which we have deep limits (F336W, F435W, and F555W), our most constraining limit comes from ACS F555W with $m_{\rm F555W}>27.1$~mag, corresponding to $M_{\rm F555W}>-2.2$~mag with no additional extinction from circumstellar matter.  This could be the case for a companion at wide separations with minimal additional extinction.  For BPASS models with Solar metallicity, this limit excludes any systems with a secondary star with $>$6.4~$M_{\odot}$.  In scenarios where the $V$-band circumstellar extinction ($A_{V}=4.4$~mag, implying $M_{\rm F555W}>-6.6$~mag) is taken into account, virtually all BPASS models are consistent with our limits.  A close binary therefore remains a possibility for SN\,2023ixf, whose presence could better be constrained with deep optical imaging after the SN fades.

\begin{figure}
    \centering
    \includegraphics[width=0.49\textwidth]{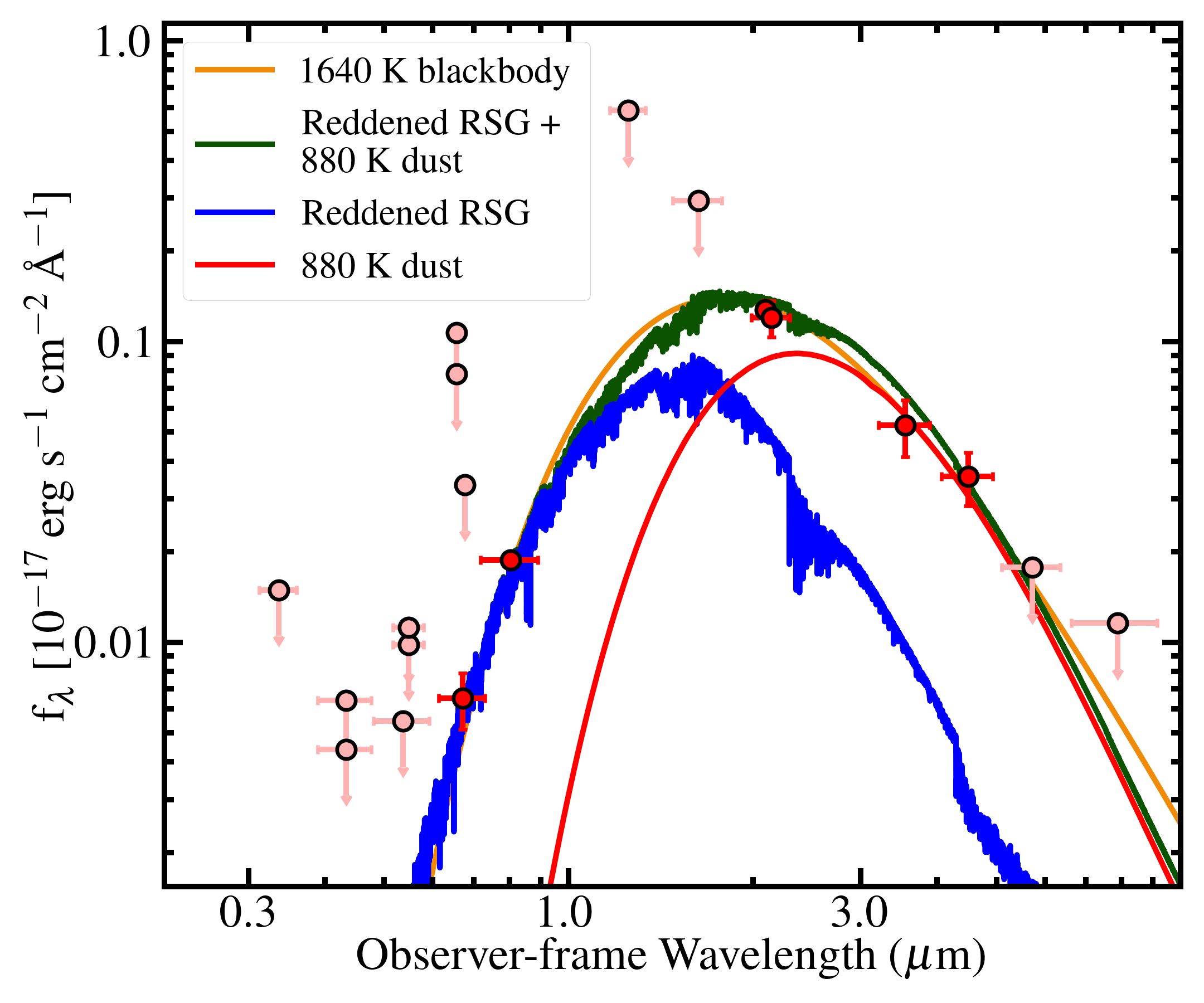}
    \caption{SED of the pre-explosion counterpart to SN\,2023ixf, with red circles denoting detections and pink circles denoting upper limits (described in \autoref{sec:sed}).  We fit the {\it HST}, {\it Spitzer}, and ground-based photometry photometry with a 1640~K blackbody (orange line), which describes the data but is much cooler than typical effective temperatures for the RSG progenitor stars of SN\,II \citep[e.g., in][]{Smartt15}.  We also show a RSG SED for a reddened RSG supergiant with a $T_{\rm eff}=3780$~K photosphere inside of a 880~K dust shell exhibiting mid-infrared excess \citep[green; from][]{Kilpatrick18:17eaw}.  The individual components of the overall reddened RSG SED (star and dust shell) are shown in blue and red, respectively}
    \label{fig:sed}
\end{figure}

\begin{figure}
    \includegraphics[width=0.49\textwidth]{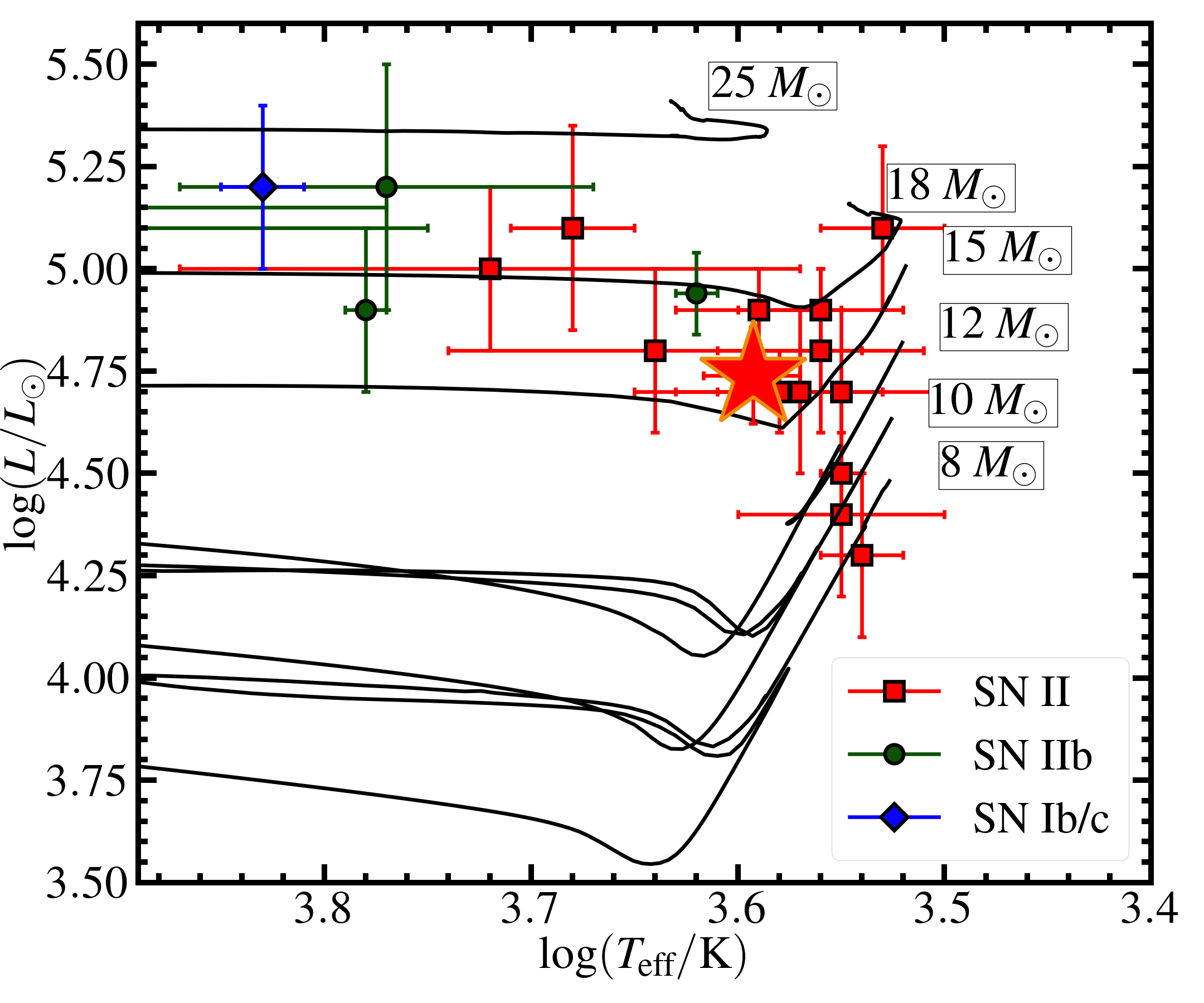}
    \caption{A Hertzsprung-Russell diagram zoomed in on the red supergiant branch.  The location of the SN\,2023ixf progenitor candidate inferred from our RSG spectral model (Section~\ref{sec:sed}) is shown as a red star.  For context, we show the locations of other SN\,II progenitor stars from \citet{Smartt15} as red squares, the progenitor stars of the SN\,IIb 1993J \citep{aldering+94}, 2011dh \citep{maund+11}, and 2013df \citep{vandyk+14} as green circles, and the progenitor candidate of the SN\,Ib 2019yvr \citep{Kilpatrick21} as a blue diamond.  The black lines are single-star evolutionary tracks from MIST \citep{choi+16} as described in Section~\ref{sec:sed}.}
\end{figure}

\subsection{Pre-Explosion Variability of the Progenitor Candidate and Implications for Mass Loss}\label{sec:variability}

The SN\,2023ixf progenitor star exhibited extreme variability in the \spitzer/IRAC bands several years before explosion (Fig.~\ref{fig:spitzer}).  We show the light curve of the counterpart at 3.6 and 4.5~$\mu$m from $\sim$2012--2020 in Fig.~\ref{fig:spitzer}.  For comparison, we also show the 3.9~$\mu$m light curve of $\alpha$ Orionis (Betelgeuse) from 1 January 2017 to 12 December 2022 \citep{Taniguchi22}, but shifted in time by 4~yr so it overlaps with that of the SN\,2023ixf counterpart and scaled to its average flux at $\sim$23~$\mu$Jy.  In the SED analysis above, we emphasize that we accounted for intrinsic variations in the {\it Spitzer} bands by including the standard deviation across the light curves in the average Channel 1 and 2 fluxes.

The SN\,2023ixf progenitor star exhibited significant mid-IR variability with an average of 22.44 and 23.99~$\mu$Jy and peak-to-peak variability of 15.64 and 17.86~$\mu$Jy at 3.6 and 4.5~$\mu$m, respectively (roughly 0.8~mag or 70\% in both bands).  These extreme variations appears correlated in the two {\it Spitzer} bands with approximately the same overall magnitude, which would be consistent with a mode of variability where the visible photosphere expands and contracts with at most small variations in effective temperature.  We also note that these variations are similar in amplitude to the high-luminosity end of large-amplitude, cool pulsators observed in the LMC \citep{OGrady20,OGrady23}.

Moreover, the light curve from 2012--2020 exhibits as quasi-sinusoidal variation with a timescale of roughly 2.8~yr (1000~day).  We infer this timescale via Fourier transform of the {\it Spitzer}/IRAC photometry, although the sampling of the light curve and the fact that we only observe peak-to-peak variations over $\sim$2.5 cycles in Fig.~\ref{fig:spitzer} suggests it is uncertain and could vary anywhere from 2.6--3.0~yr.  Combined, this evidence is similar to $\kappa$-mechanism pulsations in Betelgeuse, which are the primary mode of variability in that star and are driven by changes in the atmospheric opacity \citep[see, e.g.,][]{Li94,Heger97}.  \citet{Paxton13} observe these modes with timescales of 1--8~yr directly in {\tt MESA} models of RSGs where the structure of the star is resolved with sufficiently high time resolution, also in close agreement with simulations in \citet{Yoon10} and our inferred timescale.  These pulsations drive expansion and contraction in the atmosphere at a nearly constant temperature \citep[e.g.,][]{Levesque20}, resulting in overall changes to the luminosity from Betelgeuse.

In addition, the significant variability correlated across both bands supports the conclusion that the {\it Spitzer} counterpart is dominated by a single source.  As this variability is extreme in the IR even for a single RSG, the flux is unlikely to contain significant emission from a second source (e.g., Source B) compared with the minimum flux level of our light curve in Fig.~\ref{fig:spitzer}.

\begin{figure}
    \includegraphics[width=0.49\textwidth]{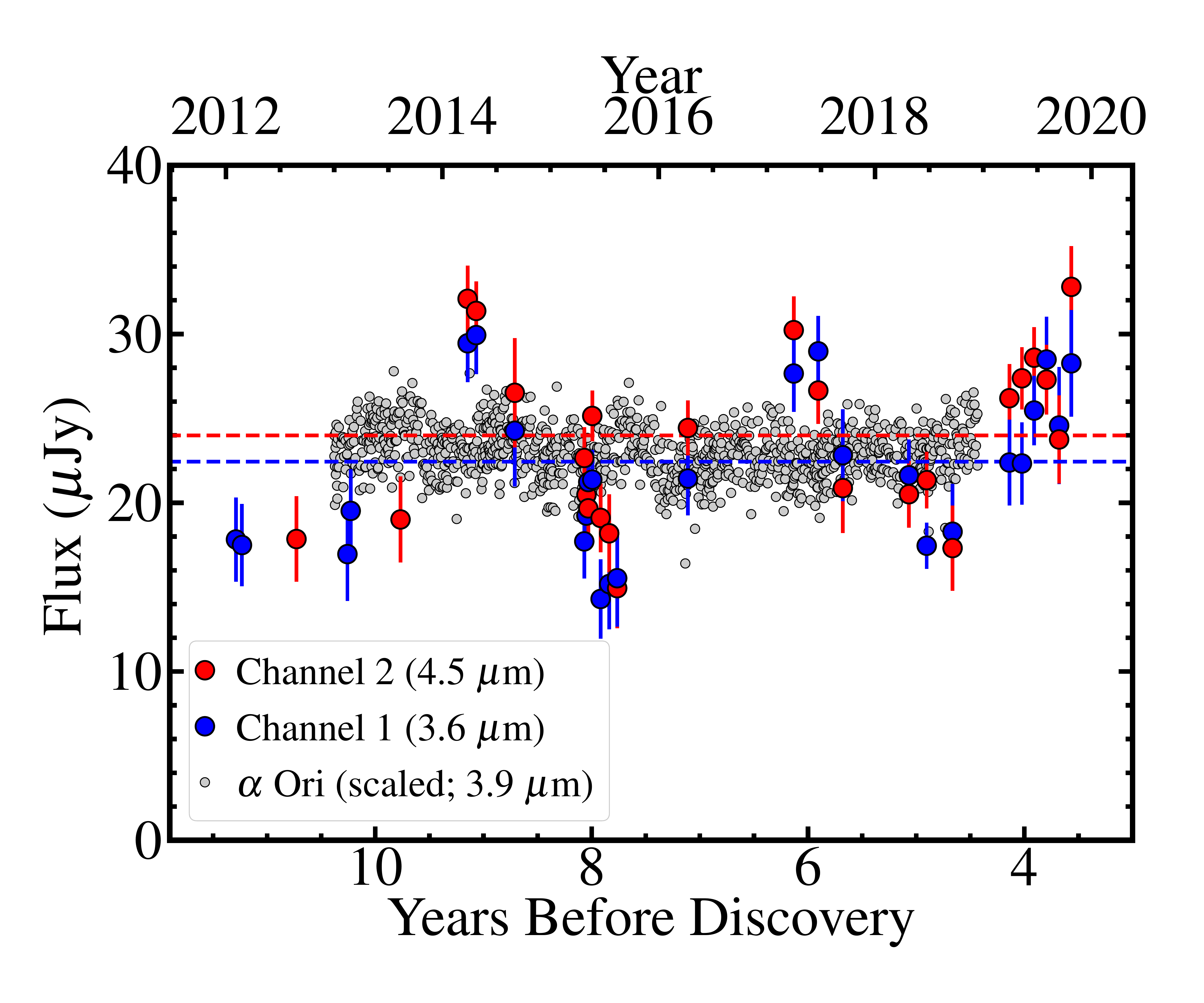}
    \caption{Light curve of the pre-explosion counterpart to SN\,2023ixf from \spitzer/IRAC Channel 1 and 2 observations from $\sim$11.3--3.6~yr prior to discovery of the supernova. We note that both channels exhibit $\sim$16~$\mu$Jy peak-to-peak variability during this time frame, with a significant peaks spaced $\sim$2.8~yr apart.  The flux-weighted average in Channels 1 and 2 are shown as blue and red dashed lines, respectively, with their values given in \autoref{tab:spitzer}.}\label{fig:spitzer}
\end{figure}

Assuming this mechanism is responsible for the variability in the SN\,2023ixf counterpart, the required changes in luminosity are $\approx$50\% larger than in Betelgeuse.  Our derived mass-loss rate is comparable with Betelgeuse \citep[which has a rate from 0.2--2$\times10^{-6}~M_{\odot}~\text{yr}^{-1}$;][]{Dolan16}, which matches expectations for the $\kappa$-mechanism driving strong mass loss with large variations in luminosity.  However, our estimate precludes a ``superwind'' generated in the CSM \citep[e.g.,][]{Yoon10,Davies22} up to the point where our data cut off 3.6~yr before explosion.  The pre-explosion mass-loss rate could be enhanced if the star was significantly more active during the final 3~yr before core collapse, which is predicted by \citet{Jacobson-Galan23}. 

\section{Discussion}\label{sec:discussion}

Current estimates on the maximum luminosity of SN\,II progenitor stars are dominated by a handful of direct counterpart detections in the literature \citep[e.g., in][]{Davies18,Kochanek20}, and the highest luminosity stars in those samples contribute significant weight to statistical analyses of the distribution from which they are drawn.

Assuming the SN\,2023ixf candidate counterpart is dominated by its progenitor star emission in {\it HST}, {\it Spitzer}, and ground-based imaging, SN\,2023ixf provides the best examples to date of the optical to mid-IR SED of a SN\,II progenitor star \citep[comparable to SN\,2017eaw;][]{Kilpatrick18:17eaw,Rui19,vanDyk19}, but with a low foreground host-galaxy extinction and precise distance.  The significant variability and large implied circumstellar extinction support the presence of such features in other SN\,II progenitor stars given the lack of multi-epoch, multi-band imaging in which they can be analyzed.  The vast majority of directly-detected SN\,II progenitor stars in recent analyses \citep{Smartt15,Davies18,Kochanek20} were identified from F814W imaging.  Our analysis of SN\,2023ixf demonstrate that it has extreme bolometric corrections when accounting for its mid-IR luminosity.  We conclude that the red supergiant problem can in part be mitigated by extreme circumstellar extinction, and systems with {\it Spitzer} or {\it JWST} detections similar to SN\,2023ixf can constrain the distribution of that extinction from their IR excess.

\section{Conclusion}\label{sec:conclusions}

We have presented direct imaging showing that there is a credible progenitor candidate to the type II SN\,2023ixf in M101 at 6.85~Mpc.  In summary, this imaging demonstrates:

\begin{enumerate}
    \item The candidate progenitor star to SN\,2023ixf is most consistent with a $\log(L/L_{\odot})=4.74\pm0.07$ RSG with an effective temperature of $T_{\rm eff}=3920\substack{+200\\-160}$~K.  Following single-star evolutionary tracks, this would place the progenitor star's initial mass at 11$\pm$2~$M_{\odot}$, placing it within the range of other low to moderate mass RSG progenitors to SN\,II \citep[e.g., in][]{Smartt15}.
    \item Modeling of the mid-IR SED from this counterpart suggests that it was enshrouded in a dusty shell of CSM similar to SN\,2017eaw \citep{Kilpatrick18:17eaw}.  The implied mass-loss rate for a wind that could produce this shell divided by its wind speed is $\dot{M}/v_{\rm wind}=1.3\pm0.1\times10^{-6}~M_{\odot}~\text{yr}^{-1}/(50~\text{km~s}^{-1})$.  This is comparable to more massive stars such as Betelgeuse but low compared to a ``superwind'' or mass-loss rates from immediately before explosion inferred in \citet{Jacobson-Galan23}.
    \item \spitzer/IRAC photometry exhibits significant pre-explosion variability that is correlated in both bands.  We also see evidence for a 2.8~yr (1000~day) timescale in this variability, similar to but generally stronger in amplitude than pulsations in other RSGs driven by opacity changes in their atmosphere \citep[i.e., the $\kappa$ mechanism][]{Li94,Heger97,Yoon10,Paxton13}.
\end{enumerate}

Future studies of the progenitor stars to SN\,II will greatly benefit from multi-band, multi-epoch imaging of resolved stellar populations using deep optical and IR surveys of nearby galaxies, such as those by the Vera C. Rubin Observatory and {\it Nancy Grace Roman Space Telescope} \citep{Ivezic19,Gezari22}.  To better understand these stars, their pre-explosion evolution, and the exact cause of the red supergiant problem, only detailed SEDs and light curves from optical to mid-IR, such as those that we present for the SN\,2023ixf counterpart, can shed light on the pathways through which SN\,II progenitor stars evolve and explode.

\bigskip\bigskip\bigskip
\noindent {\bf ACKNOWLEDGMENTS}
\smallskip

C.D.K. is partly supported by a CIERA postdoctoral fellowship and acknowledges support from a NASA grant for HST-GO-16136.
M.R.D. acknowledges support from the NSERC through grant RGPIN-2019-06186, the Canada Research Chairs Program, the Canadian Institute for Advanced Research (CIFAR), and the Dunlap Institute at the University of Toronto.
A.G. is supported by the National Science Foundation Graduate Research Fellowship Program under Grant No. DGE–1746047. A.G. also acknowledges funding from the Center for Astrophysical Surveys Fellowship at UIUC/NCSA and the Illinois Distinguished Fellowship. K.S.M. acknowledges funding from the European Union’s Horizon 2020 research and innovation programme under ERC Grant Agreement No. 101002652 and Marie Skłodowska-Curie Grant Agreement No. 873089.
C.G. is supported by a VILLUM FONDEN Young Investigator Grant (project number 25501).
J.H. was supported by a VILLUM FONDEN Investigator grant (project number 16599).
S.J.S. acknowledges funding from STFC grants ST/X006506/1 and ST/T000198/1.

The UCSC team is supported in part by NASA grants NNG17PX03C and 80NSSC22K1518, NSF grant AST--1815935, the Gordon \& Betty Moore Foundation, the Heising-Simons Foundation, and by a fellowship from the David and Lucile Packard Foundation to R.J.F.

The Young Supernova Experiment (YSE) and its research infrastructure is supported by the European Research Council under the European Union's Horizon 2020 research and innovation programme (ERC Grant Agreement 101002652, PI K.\ Mandel), the Heising-Simons Foundation (2018-0913, PI R.\ Foley; 2018-0911, PI R.\ Margutti), NASA (NNG17PX03C, PI R.\ Foley), NSF (AST-1720756, AST-1815935, PI R.\ Foley; AST-1909796, AST-1944985, PI R.\ Margutti), the David \& Lucille Packard Foundation (PI R.\ Foley), VILLUM FONDEN (project 16599, PI J.\ Hjorth), and the Center for AstroPhysical Surveys (CAPS) at the National Center for Supercomputing Applications (NCSA) and the University of Illinois Urbana-Champaign.

The Pan-STARRS1 Surveys (PS1) and the PS1 public science archive have been made possible through contributions by the Institute for Astronomy, the University of Hawaii, the Pan-STARRS Project Office, the Max-Planck Society and its participating institutes, the Max Planck Institute for Astronomy, Heidelberg and the Max Planck Institute for Extraterrestrial Physics, Garching, The Johns Hopkins University, Durham University, the University of Edinburgh, the Queen's University Belfast, the Harvard-Smithsonian Center for Astrophysics, the Las Cumbres Observatory Global Telescope Network Incorporated, the National Central University of Taiwan, STScI, NASA under grant NNX08AR22G issued through the Planetary Science Division of the NASA Science Mission Directorate, NSF grant AST-1238877, the University of Maryland, Eotvos Lorand University (ELTE), the Los Alamos National Laboratory, and the Gordon and Betty Moore Foundation.

YSE-PZ \citep{Coulter23} was developed by the UC Santa Cruz Transients Team with support from NASA grants NNG17PX03C, 80NSSC19K1386, and 80NSSC20K0953; NSF grants AST-1518052, AST-1815935, and AST-1911206; the Gordon \& Betty Moore Foundation; the Heising-Simons Foundation; a fellowship from the David and Lucile Packard Foundation to R.J.F.; Gordon and Betty Moore Foundation postdoctoral fellowships and a NASA Einstein fellowship, as administered through the NASA Hubble Fellowship program and grant HST-HF2-51462.001, to D.O.J.; and a National Science Foundation Graduate Research Fellowship, administered through grant No.\ DGE-1339067, to D.A.C.

This research is based on observations made with the NASA/ESA Hubble Space Telescope obtained from the Space Telescope Science Institute, which is operated by the Association of Universities for Research in Astronomy, Inc., under NASA contract NAS 5–26555. These observations are associated with programs 6829 (PI Chu), 9490 (PI Kuntz), 9720 (PI Chandar), and 13361 (PI Blair).
This work is based in part on archival data obtained with the Spitzer Space Telescope, which was operated by the Jet Propulsion Laboratory, California Institute of Technology under a contract with NASA. Support for this work was provided by an award issued by JPL/Caltech.
Based on observations obtained at the international Gemini Observatory, a program of NSF’s NOIRLab, which is managed by the Association of Universities for Research in Astronomy (AURA) under a cooperative agreement with the National Science Foundation on behalf of the Gemini Observatory partnership: the National Science Foundation (United States), National Research Council (Canada), Agencia Nacional de Investigaci\'{o}n y Desarrollo (Chile), Ministerio de Ciencia, Tecnolog\'{i}a e Innovaci\'{o}n (Argentina), Minist\'{e}rio da Ci\^{e}ncia, Tecnologia, Inova\c{c}\~{o}es e Comunica\c{c}\~{o}es (Brazil), and Korea Astronomy and Space Science Institute (Republic of Korea).
This work was enabled by observations made from the Gemini North telescope, located within the Maunakea Science Reserve and adjacent to the summit of Maunakea. We are grateful for the privilege of observing the Universe from a place that is unique in both its astronomical quality and its cultural significance.

\textit{Facilities}: Gemini (GMOS, NIRI), {\it HST} (WFPC2, ACS, WFC3), Mayall (NEWFIRM), {\it Spitzer} (IRAC)

\textit{Software}:
{\tt astropy} \citep{astropy},
{\tt dolphot} \citep{dolphot},
{\tt DoPhot} \citep{schechter+93},
{\tt drizzlepac} \citep{drizzlepac},
{\tt emcee} \citep{emcee},
{\tt hst123} \citep{hst123}
{\tt IRAF} \citep{IRAF},
{\tt MOPEX} \citep{Makovoz05},
{\tt pyraf} \citep{pyraf},
{\tt swarp} \citep{swarp},
YSE-PZ \citep{Coulter22, Coulter23}

\bigskip

\section*{Data and Software Availability}

All data and analysis products presented in this article are available upon request.  Analysis code and photometry used in this paper are available at \url{https://github.com/charliekilpatrick/progenitors}.  The {\it Hubble Space Telescope} data used in this paper can be found in MAST: \dataset[10.17909/dqc4-yx93]{http://dx.doi.org/10.17909/dqc4-yx93}.  The {\it Spitzer Space Telescope} data used in this paper can be found in IRSA: \dataset[10.26131/IRSA543]{http://dx.doi.org/10.26131/IRSA543}.  Gemini data are publicly available on the Gemini data archive at \url{https://archive.gemini.edu/} and NEWFIRM data are from \url{https://astroarchive.noirlab.edu/}.

\bibliographystyle{aasjournal} 
\bibliography{2023ixf}

\end{document}